\documentclass[prl,10pt,twocolumn,showkeys,
nofootinbib,
 amsmath,amssymb,
 aps,
]{revtex4-2}
\usepackage{yfonts}
\usepackage{graphicx}
\usepackage{dcolumn}
\usepackage{bm}
\usepackage{nicefrac}
\usepackage{slashed}
\usepackage{xcolor}
\definecolor{darkblue}{RGB}{0,0,196}
\definecolor{darkgreen}{RGB}{0,120,0}
\definecolor{magenta}{RGB}{255,0,255}
\usepackage[colorlinks=true,linktocpage=true,linkcolor=darkblue,citecolor=red,urlcolor=darkblue]{hyperref}
\usepackage{hyperref}
\usepackage{cancel}

\newcommand{\beq}{\begin{equation}}
\newcommand{\eeq}{\end{equation}}

\newcommand{\bea}{\begin{eqnarray}}
\newcommand{\eea}{\end{eqnarray}}

\newcommand{\bel}[1]{\begin{eqnarray}\label{#1}}
\newcommand{\eel}{\end{eqnarray}}

\def\LB{\left(}
\def\RB{\right)}

\newcommand{\EQ}[1]{Eq.~(\ref{#1})}
\newcommand{\EQn}[1]{(\ref{#1})}

\newcommand{\EQS}[1]{Eqs.~(\ref{#1})}

\newcommand{\EQSTWO}[2]{Eqs.~(\ref{#1})~and~(\ref{#2})}
\newcommand{\EQSTWOn}[2]{(\ref{#1})~and~(\ref{#2})}

\newcommand{\EQSM}[2]{Eqs.~(\ref{#1})--(\ref{#2})}
\newcommand{\EQSMn}[2]{(\ref{#1})--(\ref{#2})}

\newcommand{\CIT}[1]{Ref.~\citep{#1}} 
\newcommand{\CITn}[1]{\citep{#1}} 

\newcommand{\dd}{\mathrm{d}}

\def\epsUabgd{\epsilon^{\alpha \beta \gamma \delta}}

\newcommand{\Pv}{{\boldsymbol P}}

\newcommand{\av}{{\boldsymbol a}} 
\newcommand{\bv}{{\boldsymbol b}} 
\newcommand{\bvp}{{\boldsymbol b}^\prime} 
\newcommand{\ev}{{\boldsymbol e}}
\newcommand{\evp}{{\boldsymbol e}^\prime}

\newcommand{\nv}{{\boldsymbol n}} 

\newcommand{\vv}{{\boldsymbol v}}
\newcommand{\pv}{{\boldsymbol p}}

\newcommand{\sv}{{\boldsymbol s}}
\newcommand{\xv}{{\boldsymbol x}}

\newcommand\alphav{{\boldsymbol \alpha}}

\newcommand\sigv{{\boldsymbol \sigma}}
\newcommand\zetav{{\boldsymbol \zeta}}

\newcommand{\trt}{{\rm tr_2}}
\newcommand{\trf}{{\rm tr_4}}

\newcommand{\f}[2]{\frac{#1}{#2}}
\newcommand{\onehalf}{{\nicefrac{1}{2}}} 
 
\newcommand{\threefourths}{{\nicefrac{3}{4}}} 

\def\spin{\,\textgoth{s:}}

\def\ubarrp{{\bar u}_r(p)}
\def\ubarsp{{\bar u}_s(p)}
\def\usp{u_s(p)}
\def\urp{u_r(p)}

\def\vbarrp{{\bar v}_r(p)}
\def\vbarsp{{\bar v}_s(p)}
\def\vsp{v_s(p)}
\def\vrp{v_r(p)}

\def\fplusrsxp{f^+_{rs}(x,p)}

\def\fminusrsxp{f^-_{rs}(x,p)}

\def\omnL{\omega_{\mu\nu}}

\def\omnLD{{\tilde \omega}_{\mu\nu}}

\begin{document}


\title{Local equilibrium Wigner function for spin-$\nicefrac{1}{2}$ particles}

\author{Samapan Bhadury}
\affiliation{Institute of Theoretical Physics, Jagiellonian University, PL-30-348 Krak\'ow, Poland}

\author{Zbigniew Drogosz}
\affiliation{Institute of Theoretical Physics, Jagiellonian University, PL-30-348 Krak\'ow, Poland}

\author{Wojciech Florkowski}
\affiliation{Institute of Theoretical Physics, Jagiellonian University, PL-30-348 Krak\'ow, Poland}

\author{Sudip Kumar Kar}
\affiliation{Institute of Theoretical Physics, Jagiellonian University, PL-30-348 Krak\'ow, Poland}

\author{Valeriya Mykhaylova}
\affiliation{Institute of Theoretical Physics, Jagiellonian University, PL-30-348 Krak\'ow, Poland}

\date{\today}
            
\begin{abstract}
Formal connections between the spin density matrix and the Wigner function for spin-$\onehalf$ particles forming a relativistic gas are explored to determine their general structures. They suggest that the commonly used form of the local equilibrium Wigner function should be replaced by a new expression. The latter fulfills the necessary condition for the normalization of the mean spin polarization, which the former fails to reproduce. The new definition of the Wigner function leads to generalized thermodynamic relations for perfect spin hydrodynamics, identical to those obtained earlier using the classical concept of spin. Moreover, one can prove that the perfect spin hydrodynamics based on the new equilibrium Wigner function is nonlinearly causal and stable. Finally, the selection rule for the Lagrange multipliers, which is satisfied by real systems, is discussed. 
\end{abstract}

                              
\maketitle

{\it Introduction ---} There is a broad interest in phenomena related to spin polarization observed in relativistic heavy-ion collisions~\CITn{Liang:2004ph, Liang:2004xn, STAR:2017ckg, STAR:2018gyt, STAR:2019erd, ALICE:2019aid}; for a recent experimental review, see~\CITn{Niida:2024ntm}. On the theory side, the polarization effects triggered the development of relativistic spin hydrodynamics~\CITn{Becattini:2009wh, Becattini:2011zz, Montenegro:2017rbu, Florkowski:2017ruc, Florkowski:2017dyn, Florkowski:2018ahw, Hattori:2019lfp, Weickgenannt:2019dks, Hattori:2019lfp, Florkowski:2019qdp, Bhadury:2020puc, Montenegro:2020paq, Weickgenannt:2020aaf, Shi:2020htn, Bhadury:2020cop, Fukushima:2020ucl, Li:2020eon, Singh:2020rht, Gallegos:2021bzp, Weickgenannt:2021cuo, She:2021lhe, Hongo:2021ona, Hu:2021pwh, Singh:2022ltu, Weickgenannt:2022zxs, Wagner:2022amr, Dey:2023hft, Weickgenannt:2023nge, Kumar:2023ojl, Wagner:2024fhf, Wagner:2024fry, Dey:2024cwo, She:2024rnx, Huang:2024ffg, Bhadury:2025fil, Dey:2025wqw}. This approach incorporates spin degrees of freedom into the well-established framework of relativistic hydrodynamics that forms the basic tool to interpret heavy-ion data~\CITn{Ollitrault:2007du, Romatschke:2009im, Florkowski:2010zz, Gale:2013da, Jaiswal:2016hex}.

Currently, spin hydrodynamics represents the forefront of theoretical investigations related to relativistic heavy-ion physics and quark-gluon plasma. Only very recently have realistic 3D hydrodynamic simulations been performed~\CITn{Singh:2024cub, Sapna:2025yss}, where spin degrees of freedom follow their own dynamic equations coupled to an ordinary hydrodynamic background. 

Many formulations of spin hydrodynamics are rooted in the concept of a local equilibrium Wigner function, originally introduced in~\CITn{Becattini:2013fla}. The form proposed in~\CITn{Becattini:2013fla} was examined in more detail in~\CITn{Florkowski:2017dyn}, where it was found to violate the normalization of the mean polarization vector that should be confined to the range between 0 and~1 (0~corresponds to no polarization, 1~describes a pure state, and all other values describe mixed states). Consequently, the form of local equilibrium Wigner function introduced in~\CITn{Becattini:2013fla} has been considered valid only for small values of the spin polarization tensor $\omega_{\alpha\beta}$ that can be interpreted as the dimensionless ratio of the spin chemical potential $\Omega_{\alpha\beta}$ to the temperature~$T$, $\omega_{\alpha\beta} = \Omega_{\alpha\beta}/T$.

Problems with local equilibrium Wigner function inspired frameworks that used the classical description of spin~\CITn{Florkowski:2018fap}. Such an approach has turned out to be especially convenient for the incorporation of nonlocal effects responsible for the dissipative spin-orbit interaction that changes the spin part of the angular momentum into the orbital part and vice versa~\CITn{Weickgenannt:2020aaf, Wagner:2022amr, Weickgenannt:2024ibf}.

In this work, we analyze the formal connections between the spin density matrix and the Wigner function for spin-$\onehalf$ particles and determine their general structures. Our results demonstrate that the commonly used form of the local equilibrium Wigner function should be replaced by a new expression that correctly defines the mean polarization vector.  Our solution eventually becomes consistent with the expression derived in the seminal Landau-Lifshitz (LL) course~\CITn{Berestetskii:1982qgu}, indicating the possibility of the application in other theories~\CITn{PhysRevLett.118.226601, PhysRevLett.87.187202}.

With the spin degrees of freedom described by the new spin density matrix (or the Wigner function), we successfully introduce the thermodynamic relations with spin, whose generic forms have been derived earlier \mbox{in~\CITn{Florkowski:2024bfw, Drogosz:2024gzv}}. In particular, we find that the results for the conserved tensors obtained with classical and quantum descriptions agree exactly to the second order in $\omega$. In this case, the integrals over allowed classical spin configurations are equivalent to the calculation of the traces over the spinor space. At higher orders of $\omega$, there are differences between the classical and quantum approaches since the classical magnitude of polarization is limited \mbox{by~$\sqrt{\threefourths}$}, while the quantum one by $\onehalf$. This behavior is studied in more detail in a separate work~\cite{Drogosz:2025iyr}. 

We additionally demonstrate that the perfect spin hydrodynamics based on the new definition of the Wigner function is nonlinearly causal and stable, which is crucial for a consistent formulation of hydrodynamic framework and its numerical applications.  In this way, we generalize the results obtained in~\cite{Abboud:2025qtg}, where this property was established for the formalism using the classical spin description. This very nontrivial feature strongly supports the form advocated in this Letter.

Finally, we discuss a selection rule for the Lagrange multipliers appearing in our formalism, which applies to real systems, i.e., those that have finite densities of energy, linear and orbital momentum, and spin. An extended discussion of this issue has been recently published in~\cite{Drogosz:2025ihp}.

To conclude, we solve herein a long-standing problem of the proper formulation of the local equilibrium Wigner function for particles with spin~$\onehalf$. 

\medskip
{\it Notation and conventions ---} For the Levi-Civita symbol $\epsilon^{\mu\nu\alpha\beta}$, we follow the convention \mbox{$\epsilon^{0123} =-\epsilon_{0123} = +1$}. The tensor  \mbox{${\tilde a}^{\alpha\beta} \equiv (\onehalf) \,\epsilon^{\alpha\beta\gamma\delta} a_{\gamma \delta}$} is dual to $a^{\alpha\beta}$. The metric tensor is of the form \mbox{$g_{\mu\nu} = \textrm{diag}(+1,-1,-1,-1)$}. The trace over spinor (spin) indices is denoted by $\trf$ ($\trt$). Throughout the text, we make use of natural units, \mbox{$\hbar = c = k_{\rm B} = 1$}. The scalar product of two four-vectors $a$ and $b$ reads $a \cdot b = a^0 b^0 - \av \cdot \bv$, where the three-vectors are indicated in bold. The components of the four-momentum of a particle with mass $m$ are \mbox{$p^\mu = (E_{\rm p},\pv)$}, with \mbox{$E_{\rm p} =\sqrt{m^2 + \pv^2}$}. 

\medskip
{\it Spin and spinor density matrices ---} We consider \mbox{spin-$\nicefrac{1}{2}$} particles and antiparticles. Their spin can be described by two-by-two spin density matrices \mbox{$f^\pm(x,p)$} or by four-by-four spinor density matrices $X^\pm(x,p)$, defined for each value of the space-time position $x^\mu = (t,\xv)$ and the particle four-momentum $p^\mu = (E_{\rm p}, \pv)$.\footnote{We use the names ``spin density'' and ``spinor density'' (matrix) to easily distinguish between the functions $f^\pm(x,p)$ and $X^\pm(x,p)$. In the literature, often one and the same name is used for these two objects (spin density).}~They are related through the equations:
\bel{fplusrsxp}
\left[ f^+(x,p) \right]_{rs}  \equiv  \fplusrsxp &=&  \frac{1}{2m} \,\ubarrp X^+ \usp,  \\
\left[ f^-(x,p) \right]_{rs}  \equiv \fminusrsxp &=& - \frac{1}{2m} \, \vbarsp X^- \vrp,
\label{fminusrsxp} 
\eel
where $\urp$ and $\vsp$ are Dirac spinors (with spin indices $r$ and $s$ running from 1~to~2), while the bar denotes their Dirac adjoints. We use the normalization conditions $\ubarrp \usp=2m\,\delta_{rs}$ and \mbox{$\vbarrp \vsp=-2m\,\delta_{rs}$.} For~more details on our conventions for $\urp$ and $\vsp$, see the Supplemental Material.

\smallskip
{\it Equilibrium spinor density (current status) ---} Following the original idea formulated in~\cite{Becattini:2013fla}, a common prescription currently used in the literature for $X^\pm$ in local equilibrium is 
\bel{XpmM}
X^{\pm} =  \exp\left[\pm \xi(x) - \beta_\mu(x) p^\mu \right] M^\pm ,
\eel
with
\bel{Mpm}
M^\pm = \exp\left[ \pm \f{1}{2} \omega_{\mu\nu}(x)  \Sigma^{\mu\nu} \right] .
\eel
In \EQ{XpmM}, $\beta^\mu= u^\mu/T$, $u^\mu = \gamma (1, \vv )$ is the fluid four-velocity (normalized to unity), and $\xi = \mu/T$, where $\mu$ is the baryon chemical potential. Further, the matrix $\Sigma^{\mu\nu}$ in \EQ{Mpm} is defined as \mbox{$\Sigma^{\mu\nu}=\sigma^{ \mu\nu}/2=(i/4)[\gamma^\mu,\gamma^\nu]$}, and the quantity $\omega_{\mu\nu}$ is the spin polarization tensor, which can be expressed in terms of the electric- and magnetic-like three-vectors, \mbox{$\ev = (e^1,e^2,e^3)$} and $\bv = (b^1,b^2,b^3)$, as
\bel{omeb}
\omega_{\mu\nu} = 
\begin{bmatrix}
0       &  e^1 & e^2 & e^3 \\
-e^1  &  0    & -b^3 & b^2 \\
-e^2  &  b^3 & 0 & -b^1 \\
-e^3  & -b^2 & b^1 & 0
\end{bmatrix},
\eel
where the sign conventions follow those of~\cite{Jackson:1998nia}.

It has been shown in \CIT{Florkowski:2017dyn} that the mean spin polarization for both particles and antiparticles in equilibrium is defined by the equation
\beq
\Pv =  \f{1}{2} \f{ \trt \left( f^\pm \sigv\right)  }{\trt \left( f^\pm \right) } 
\eeq
and equals
\begin{eqnarray}
\hspace{-0.5cm} \Pv &=& -\f{1}{2} \tanh\left(  
\frac{ \sqrt{ \bv_\ast \cdot \bv_\ast\!-\! \ev_\ast \cdot \ev_\ast }}{2} \right) 
\frac{ \bv_\ast }{\sqrt{ \bv_\ast \cdot \bv_\ast \!-\! \ev_\ast \cdot \ev_\ast }},
\label{eq:problem}
\end{eqnarray}
where $\sigv$ denotes a three-vector consisting of the Pauli matrices, and the asterisks denote the values of the three-vectors $\ev$ and $\bv$ in the particle rest frame (PRF). Unfortunately, \EQ{eq:problem} is not completely satisfactory, as the proposed formalism does not guarantee that \mbox{$\bv_\ast \cdot \bv_\ast \!- \ev_\ast \cdot \ev_\ast > 0$,} and, moreover, the vector \mbox{$\bv_\ast / \sqrt{ \bv_\ast \cdot \bv_\ast \!-\! \ev_\ast \cdot \ev_\ast }$} cannot be interpreted as a unit vector that defines the direction of the mean polarization.\footnote{Note that $(1/2) \, \omega_{\mu\nu}  \omega^{\mu\nu} =   \bv \cdot \bv - \ev \cdot \ev =  \bv_\ast \cdot \bv_\ast\!-\! \ev_\ast \cdot \ev_\ast.$}  Consequently, \EQ{eq:problem} has been used for small values of $\ev$ and $\bv$, in which case it gives
\begin{eqnarray}
\Pv = -\frac{\bv_\ast(x,\pv)}{4} .
\label{eq:smallP}
\end{eqnarray}
However, we note that the smallness of $\ev$ and $\bv$ does not imply a small $\bv_\ast$, since $\bv_\ast$ depends on the momentum of a particle that might be very large. For a discussion of this point, see the Supplemental Material.

\medskip
{\it From spin to spinor density ---} As the methodology outlined above leads to problems with the correct definition of the mean spin polarization, one may try to reverse the line of the arguments. We may start with the general forms of the two-by-two spin density matrices and derive, using \EQSTWO{fplusrsxp}{fminusrsxp}, the corresponding forms of $X^\pm$.

Since $f^+_{rs}(x,p)$ and $f^-_{rs}(x,p)$ are Hermitian matrices, we can use the decomposition~\CITn{Florkowski:2017dyn}
\bel{fpm}
f_{rs}^\pm(x,p) = f^\pm_0(x,p) \left[\delta_{rs}  + \zetav_\ast^\pm(x,\pv) \cdot \sigv_{rs} \right],
\eel
where $f^\pm_0(x,p)$ is the phase-space density of (anti)particles averaged over spin.
The three-vector $\zetav_\ast^\pm(x,\pv)$ can be interpreted as a spatial part of the polarization four-vector $\zeta^{\pm \mu}_\ast(x,\pv)$ with a vanishing zeroth component, 
\begin{equation}\label{eq:zetastar}
\zeta^{\pm \mu}_\ast = \left(0, \zetav_\ast^\pm \right) .
\end{equation}
As discussed above, the magnitude of the vector $\zetav_\ast^\pm$ is restricted to the range $0 \leq |\zetav_\ast^\pm| \leq 1$. In the LAB frame, where particles move with momentum $\pv$, the form of $\zeta^{\pm \mu}$ is
\bel{eq:Zetamu}
\hspace{-0.5cm} \zeta^\mu_\pm = \Lambda^\mu_{\,\,\,\nu}(\vv_p) \zeta^\nu_{\pm \ast} = \left(\frac{\pv \cdot \zetav^\pm_\ast}{m} ,\ \zetav^\pm_\ast + \frac{\pv \cdot \zetav^\pm_\ast}{m (E_{\rm p} + m)} \pv \right),
\eel
where $\Lambda^\mu_{\,\,\,\nu}(\vv_p)$ is the canonical boost depending on the particle three-velocity $\vv_\pv = \pv/E_{\rm p}$~\CITn{Florkowski:2017dyn}. By referring to Lorentz covariance or by making a direct calculation, one can verify the relations:
\bel{mainplus}
\ubarrp \gamma_5 \zeta^+_\mu \gamma^\mu \usp = 2m \,\zetav^+_* \cdot \sigv_{rs},
\eel
and
\bel{mainminus}
\vbarsp \gamma_5 \zeta^-_\mu \gamma^\mu \vrp = -2m \,\zetav^-_* \cdot \sigv_{rs} .
\eel
This immediately indicates that
\bel{XpmNEW}
X^{\pm}_s(x,p) &=& f_0^\pm(x,p) \left[ 1 + \gamma_5  \slashed{\zeta}^\pm(x,p) \right].
\eel
Here we used the Feynman slash notation, \mbox{$\slashed{\zeta} = \zeta_\mu \gamma^\mu$}, and added a subscript $s$ to distinguish the last formula from the previous definition of $X^\pm$. In fact, we observe that \EQ{XpmNEW} disagrees with \EQ{XpmM}. This difference is particularly striking since \EQ{XpmNEW} is general and does not refer to any equilibrium properties of the spin density matrix. 

However, there is the possibility that our new formula for $X^\pm_s$ delivers the same results as the traditional expressions. It is so if we define \mbox{$\zetav^\pm_\ast(x,\pv) = -\frac{1}{2} \bv_\ast(x,\pv)$}. We must bear in mind that this agreement may hold only for very small values of $\bv_\ast(x,\pv)$ and $\ev_\ast(x,\pv)$. This finding was already discussed in~\CITn{Florkowski:2019gio}; however, without drawing any conclusions. 

To proceed further, it is convenient to make use of the identity
\begin{equation}\label{eq:expa}
\exp\!\left( \gamma_5 \slashed{a}_\pm \right)\!\!=\!\cosh\!\sqrt{-a^2_\pm}
\!\left[\!1+\!\frac{\gamma_5 \slashed{a}_\pm}{\sqrt{-a^2_\pm}} \tanh\!\sqrt{-a^2_\pm} \right]\!\!. 
\end{equation}
Here $a^\mu_\pm$ are spacelike four-vectors satisfying the condition $a^2_\pm < 0$. The expressions in the square brackets in Eqs.~\EQn{XpmNEW} and \EQn{eq:expa} are equal if 
\beq
\zeta^\mu_\pm = \frac{a^\mu_\pm}{\sqrt{-a^2_\pm}} 
\tanh\sqrt{-a^2_\pm}.
\eeq
The normalization condition gives
\beq
-\zeta^2_\pm = \zetav_{\pm *}^2 = \tanh^2\sqrt{-a^2_\pm}\leq 1,
\eeq
which agrees with the normalization condition for the three-vector~$\zetav_*$ discussed below~\EQ{eq:zetastar}. For small $a^\mu_\pm$ we have $\zeta^\mu_\pm = a^\mu_\pm$. Combining \EQn{XpmNEW} and \EQn{eq:expa}, we can write
\bel{XpmNEW2}
X^{\pm}_s(x,p) &=& f_0^\pm(x,p) \exp\left( \gamma_5 \slashed{a}_\pm \right),
\eel
where  $\cosh\sqrt{-a^2_\pm}$ was absorbed into the function~$f_0^\pm(x,p)$.

\medskip
{\it Revised equilibrium spinor density ---} Our arguments leading to the form~\EQn{XpmNEW2} are based solely on the two facts: the two-by-two spin density matrix is Hermitian, and the polarization vector $\zetav_\ast^\pm$ satisfies the condition \mbox{$0 \leq |\zetav_\ast^\pm| \leq 1$}. However, the exponential form of~\EQn{XpmNEW2} directly suggests that it can be used as an equilibrium spin density, especially in the case of the Boltzmann statistics, in which we define
\bel{XpmNEW3}
X^{\pm}_s(x,p) =  \exp\left[\pm \xi(x) - \beta_\mu(x) p^\mu + \gamma_5 \slashed{a} \right].
\eel
Above we have assumed that the vector $a$ can be interpreted as the vector Lagrange multiplier that is the same for particles and antiparticles. Furthermore, since spin thermodynamics requires, in fact, six independent Lagrange multipliers, it is convenient to introduce the relation
\bel{eq:amu}
a_\mu(x,p) = -\frac{1}{2 m} {\tilde \omega}_{\mu\nu}(x)p^\nu,
\eel
with the dual polarization tensor ${\tilde \omega}_{\mu\nu}$.\footnote{The choice of the dual spin polarization tensor in~\EQn{eq:amu} can be checked a posteriori to be consistent with other frameworks of spin hydrodynamics, for example, based on the classical spin method.} The orthogonality condition  $a_\mu(x,p) p^\mu = 0$  may be treated as our version of the Frenkel condition that is frequently used in spin hydrodynamics with different formulations~\CITn{Frenkel:1926zz}.

\medskip
{\it Wigner function ---} Using our expression for the spinor density matrix, we can calculate the Wigner function. With the definitions of the Wigner functions \mbox{$W^\pm(x,k)$} introduced in~\CITn{deGroot:1980dk}, we find 
\bel{eq:Wpm}
W^\pm(x,k)\!=\!\frac{1}{4 m}  \int \!\dd P\,
\delta^{(4)}(k\!\mp\!p) 
(\slashed{p}\!\pm\! m) X_s^\pm
(\slashed{p}\!\pm\! m), 
\eel
where $\dd P=\dd^3p/((2\pi)^3 E_{\rm p})$. Since $[\gamma_5 \slashed{a}, \slashed{p}] =0$, we can also write
\bel{eq:Wpm2} W^\pm(x,k) = \pm \frac{1}{2}  \int \dd P\,
\delta^{(4)}(k \mp p) (\slashed{p} \pm m) X_s^\pm.
\eel
It is interesting to note that the expressions $ (\slashed{p} \pm m) X_s^\pm$ are consistent with the expression derived in the LL course for the spinor density matrix~\CITn{Berestetskii:1982qgu}, as shown in~\CITn{Florkowski:2019gio}.

Starting from the Wigner function, we can derive (see~\CITn{deGroot:1980dk}) the following expressions for the baryon current, energy-momentum tensor, and the spin tensor, respectively,
\bel{eq:Nmu} 
N^\mu(x) = \sum_{r=1}^2 \int \dd P  \, p^\mu \left[f^+_{rr}(x,p)-f^-_{rr}(x,p) \right],
\eel
\bel{eq:Tmunu} 
T^{\mu\nu}(x) = \sum_{r=1}^2 \int \dd P \, p^\mu p^\nu \left[f^+_{rr}(x,p)+f^-_{rr}(x,p) \right],
\eel
\begin{align}\begin{split}\label{eq:Slmunu} 
S^{\lambda, \mu\nu}(x) &= \f{1}{2}\sum_{r,s=1}^2 \int \dd P \, p^\lambda  \left[\sigma^{+ \mu\nu}_{sr}(p) f^+_{rs}(x,p) \right. \\
 & \hspace{1cm} \left.   + \, \sigma^{- \mu\nu}_{sr} (p) f^-_{rs}(x,p) \right],  
\end{split}\end{align}
with the matrices $\sigma^{+ \mu\nu}_{sr} (p) =1/(2m)\, \ubarsp \sigma^{\mu\nu} \urp$ and \mbox{$\sigma^{- \mu\nu}_{sr} (p) =1/(2m)\, \vbarrp \sigma^{\mu\nu} \vsp$ \cite{deGroot:1980dk,Itzykson:1980rh}.} It is also convenient to define the particle current
\bel{eq:calNmu} 
{\cal N}^\mu(x) = \sum_{r=1}^2 \int \dd P  \, p^\mu \left[f^+_{rr}(x,p) + f^-_{rr}(x,p) \right].
\eel
In the spinless case, we find ${\cal N}^\mu = P \beta^\mu$. We note that because of the pseudogauge ambiguity, different forms of conserved tensors are used in the literature. The forms introduced above are known as the GLW versions, after the surnames of the authors of~\CITn{deGroot:1980dk}. Since the GLW energy-momentum tensor is symmetric, the corresponding spin tensor is conserved as well, \mbox{$\partial_\lambda S^{\lambda, \mu\nu} = 0$}, which is not necessarily the case if other pseudogauge is \mbox{used~\cite{Florkowski:2018fap}.}

\medskip
{\it Thermodynamic relations ---} To provide the thermodynamic relations, in addition to the expressions given above, we need to introduce the entropy current. The~latter can be defined by the formula originally presented in~\CITn{Florkowski:2017ruc} 
\begin{align}\begin{split}\label{eq:Smu} 
S^\mu(x) &= -\f{1}{2} \int \dd P  \, p^\mu \left\{
\trf \, \left[ X_s^+ \, \left(\ln X_s^+ - 1\right) \right]\right. \\ 
& \left.  \hspace{1cm} + \, 
\trf \, \left[X_s^- \,  \left(\ln X_s^- - 1\right) \right]
\right\}.
\end{split}\end{align}
This leads to the relation
\bel{eq:H}
S^\mu =  T^{\mu \alpha} \beta_\alpha-\f{1}{2} \omega_{\alpha\beta} S^{\mu, \alpha \beta}
-\xi N^\mu + {\cal N}^\mu,
\eel
which in the spinless case is treated as a consequence of the extensivity of the thermodynamic variables such as entropy, energy, and baryon number. We note that an alternative way to introduce the entropy current is to use \EQ{eq:H} as the definition of $S^\mu$, since all
the expressions on the right-hand side are known. In this case, we may argue that \EQSTWO{eq:Smu}{eq:H} are simply consistent. 

Using \EQSM{eq:Nmu}{eq:calNmu} we can also derive the identity
\bel{eq:GD}
\dd{\cal N}^\mu = N^\mu \dd\xi - T^{\lambda\mu} \dd\beta_\lambda + {\scriptstyle{ \frac{1}{2} }}
 S^{\mu, \alpha \beta} \dd\omega_{\alpha\beta}. 
\eel
In the spinless case, \EQ{eq:GD} is reduced to the Gibbs-Duhem relation. Equations \EQSTWOn{eq:H}{eq:GD} lead to the first law of thermodynamics in the form
\bel{eq:firstlaw}
\dd S^\mu = - \xi \dd N^\mu + \beta_\lambda \dd T^{\lambda\mu}
- {\scriptstyle{ \frac{1}{2} }}
\omega_{\alpha\beta} \dd S^{\mu, \alpha \beta}.
\eel
The last equation is of crucial importance since it indicates that the entropy is conserved owing to the conservation of the baryon number, energy, linear momentum, and the spin part of the total angular momentum. Thus, our concept of local equilibrium assumes that the spin and orbital parts of the total angular momentum \mbox{($S$ and $L$)} are conserved separately. Spin-orbit coupling leading to a transfer between $S$ and $L$ can be included by incorporation of dissipation, as described in more \mbox{detail in~\CITn{Florkowski:2024bfw,Drogosz:2024gzv,Florkowski:2024cif}.}

Equations \EQSMn{eq:H}{eq:firstlaw} were introduced for the first time~in~\CITn{Florkowski:2024bfw} as generalized thermodynamic relations for perfect spin hydrodynamics. Their structure was implied by the kinetic theory results utilizing a classical spin description. In this work, we confirm the form of such thermodynamic relations in an approach that treats spin quantum mechanically through the properly constructed spin density matrix. We emphasize that \EQSM{eq:H}{eq:firstlaw} are valid for any values of the spin polarization \mbox{tensor $\omega$}  as long as the considered integrals exist (the point discussed below).

\medskip
{\it Classical description of spin ---} The problems discussed below \EQ{eq:problem} initiated the approach based on the classical description of spin. In this framework, motivated by the seminal paper of Mathisson from 1937~\CITn{Mathisson:1937zz,2010GReGr..42.1011M}, one introduces the spin four-vector $s$ that satisfies the two conditions: $s \cdot p = 0$ and \mbox{$s^2 = - \spin^2$}, where \mbox{$\spin^2 = 3/4$} is the eigenvalue of the Casimir operator for SU(2). \mbox{With~$s$} included as an additional variable, the physical phase space expands, and one introduces the phase-space distribution functions $f^\pm(x,p,s)$ depending on $x$, $p$, and $s$. The macroscopic currents are then obtained as integrals over momentum and spin spaces with the appropriate measures (details are provided in the Supplemental Material).

The classical-spin description eliminates the problems with the normalization of the mean polarization. Moreover, in the limit where the spin polarization tensor is small, it reproduces the result for the spin tensor obtained with \EQSTWO{XpmM}{Mpm}. Consequently, for a long time, the classical-spin approach has been the main tool to develop relativistic kinetic theory for particles with spin $\onehalf$, especially in the approximation where the spin feedback on the baryon current and the energy-momentum tensor is neglected.

Of particular interest is the comparison of the equilibrium distribution functions. In the classical-spin approach we have
\bel{eq:classEQ}
f^\pm(x,p,s) = \exp\left(\pm \xi - p \cdot \beta + \frac{1}{2} \omega_{\alpha \beta} s^{\alpha \beta} \right),
\eel
with $ s^{\alpha \beta} = (1/m) \epsilon^{\alpha\beta\gamma\delta} p_\gamma s_\delta$, while in the quantum case we use \EQ{XpmNEW3}. This suggests the correspondence between the classical and quantum descriptions,
\begin{eqnarray}
s^\mu \quad \longleftrightarrow \quad \f{1}{2} \gamma_5 \gamma^\mu.
\end{eqnarray}

Although the form of the generalized thermodynamic relations is the same as that originally found in~\CITn{Florkowski:2024bfw}, a question can be asked about the particular form of the spin corrections to $N^\mu$ and $T^{\mu\nu}$. The explicit calculations shown in the Supplemental Material indicate that the quadratic corrections to $N^\mu$ and $T^{\mu\nu}$ found in the present quantum framework are identical to those obtained within the classical-spin approach~\CITn{Florkowski:2024bfw}. This shows an intriguing equivalence of performing the integration over $s$ with the calculation of the traces over the spinor space. However,  differences appear for large values of $\bv_\ast(x,\pv)$, as in the classical case the maximum spin polarization equals $\sqrt{\nicefrac{3}{4}}$, while in the quantum case $\onehalf$, see also~\CITn{Drogosz:2025iyr}.

\medskip
{\it Nonlinear causality of spin hydrodynamics ---} Following the methods outlined in~\cite{Abboud:2025qtg}, see also~\CITn{LIU1986191, Geroch:1990bw, Peralta-Ramos:2009srp, Peralta-Ramos:2010qdp, Gavassino:2022roi}, we may show that the perfect spin hydrodynamics constructed with the new Wigner function is nonlinearly causal and its equations are symmetric hyperbolic. To prove this property we first find that the macroscopic currents can be written as derivatives of a generating \mbox{function~$\chi$}, namely: \mbox{$N^\lambda = -\partial^2 \chi/ (\partial \beta_\lambda \partial \xi)$}, \mbox{$T^{\lambda\mu} = \partial^2 \chi/ (\partial \beta_\lambda \partial \beta_\mu)$}, and $S^{\lambda, \mu\nu} = -\partial^2 \chi/ (\partial \beta_\lambda \partial \omega_{\mu\nu})$, where
\begin{equation}
\chi = \frac{1}{2} \trf \left[ 
\int \dd P \cosh\xi \left(f_0^+ + f_0^- \right) 
\exp\left( \gamma_5 \slashed{a}_\pm \right)
\right]  .
\label{eq:chi}
\end{equation}
By grouping the macroscopic currents and the Lagrange multipliers into multicomponent vectors: \mbox{$N^{\lambda A} = (N^\lambda, T^{\lambda \mu}, S^{\lambda, \mu\nu}) $} and \mbox{$\zeta^B = (\xi, -\beta^\mu, \omega^{\mu\nu})$}, we construct a test quantity
\begin{equation}
M^{\lambda A B}(\zeta) \equiv  \left( \frac{\partial N^{\lambda A}}{\partial \zeta_B} \right)
= - \left( \frac{\partial^3 \chi}{\partial \zeta_B \partial \zeta_A \partial \beta_\lambda} \right).
\label{eq:MlAB}
\end{equation}
It turns out that the spin hydrodynamic theory based on the conservation laws
\begin{equation}
\partial_\lambda N^{\lambda A} = 0 \quad \Leftrightarrow \quad
M^{\lambda A B} \partial_\lambda \zeta_B = 0
\end{equation}
is nonlinearly causal if the vector \mbox{$M^{\lambda A B}(\zeta) Z_A Z_B $} is timelike and future-oriented for any nonvanishing real vector $Z$~\cite{Abboud:2025qtg}. Starting directly from (\ref{eq:MlAB}), with $\chi$ defined \mbox{by~(\ref{eq:chi})}, one can prove that 
\begin{equation}
M^{\lambda A B}(\zeta) Z_A Z_B  = \int \, \dd P \,  p^\lambda \, w(p,\zeta,Z), 
\end{equation}
where $w(p,\zeta,Z)$ is a positive function (we provide details of this calculation in the Supplemental Material). Since the four-momentum $p^\lambda$ is time-like and future-oriented, so is the quantity $M^{\lambda A B}(\zeta) Z_A Z_B$ obtained as a sum of momenta with positive weights.

It is difficult to overestimate the importance of this result. The nonlinear causality and symmetric hyperbolicity of the equations of motion of spin hydrodynamics ensure well-posedness of the initial-value problem and stability of the theory. This, in turn, implies that the constructed framework can be applied in numerical simulations to study the dynamics of spin-polarized
fluids.

\medskip 
{\it Selection criterion ---} As long as we consider real systems with finite densities of energy, linear and orbital momentum, and spin, the macroscopic currents should be given by well-defined integrals.  Since the four-vector $a_\mu$ defined by~\EQn{eq:amu} depends on the particle momentum, the integrals over the distributions~\EQn{XpmNEW3} converge for very large values of momenta if the following condition is satisfied
\begin{eqnarray}\label{eq:criterionW}
\f{1}{2} \sqrt{ {\bvp}^2 + {\evp}^2 + 2 |\evp \times \bvp|} < \f{m}{T},
\label{eq:condition}
\end{eqnarray}
where the primes denote vectors in the fluid rest frame, i.e.~$u^{\mu}=(1,0,0,0)$. For the derivation of this formula we refer to our recent paper~\cite{Drogosz:2025ihp} and the Supplemental Material. Here we only make two points: (i) The hydrodynamic parameters used in recent calculations addressing spin polarization, for example, see~\CITn{Becattini:2021iol, Fu:2021pok}, \mbox{satisfy~\EQn{eq:condition}}. (ii) One cannot take the limit $m \to 0$ in~\EQn{eq:condition}, as our formalism defines the spin polarization in the PRF that does not exist for massless particles. 

\medskip
{\it Summary ---} In this work, we have solved a long-standing problem of the proper formulation of the local equilibrium Wigner function, which accurately defines the mean polarization vector (accounting for its correct normalization). Our new result for the spin density matrix is consistent with the expression that can be found in the LL course. Our form of the local equilibrium Wigner function offers also the possibility of a quantum derivation of the generalized thermodynamic relations for spin-$\onehalf$ particles. Moreover, it can be used to define spin hydrodynamics that is nonlinearly causal and stable. Thus, the
constructed framework can be successfully utilized in numerical simulations of spin-polarized fluids.

We have established a direct relation to the classical-spin approach -- the results for the conserved tensors exactly agree up to the second order in $\omega$ since integrals over the allowed classical spin configurations are equivalent to traces over the spinor indices in this case. This clarification is crucial to establish a consistent and uniform framework of spin hydrodynamics for spin-$\onehalf$ particles.

\medskip
\begin{acknowledgments}
\textit{Acknowledgments} --- We thank Arpan Das and Radoslaw Ryblewski for critical comments and useful discussions. S. B. would like to acknowledge the support of the Faculty of Physics, Astronomy, and Applied Computer Science, Jagiellonian University via Grant No.~LM/36/BS. This work was supported in part by the National Science Centre, Poland (NCN) Grant No.~2022/47/B/ST2/01372.
\end{acknowledgments}

\newpage

\bibliography{spin-lit}

\begin{widetext}
\begin{center}
{\bf SUPPLEMENTAL MATERIAL}
\end{center}

\section{I. Dirac spinors and their normalization}

Bispinors $u_r(p)$ and $v_s(p)$ are the free Dirac spinors given by 
\begin{equation}
    u_{r}(p)=\sqrt{E_{\rm p}+m}\begin{pmatrix}
        \varphi^{(r)}\\
        \frac{\sigv \cdot \pv }{E_{\rm p}+m} \,\, \varphi^{(r)}
    \end{pmatrix},\qquad v_r(p)=\sqrt{E_{\rm p}+m}\begin{pmatrix}
        \frac{\sigv \cdot \pv} {E_{\rm p}+m}\eta^{(r)}\\
        \eta^{(r)}
    \end{pmatrix},
\end{equation}
where
\beq
    \varphi^{(1)}=\begin{pmatrix}
        1\\
        0
    \end{pmatrix},\qquad \varphi^{(2)}=\begin{pmatrix}
        0\\
        1
    \end{pmatrix},\qquad \eta^{(1)} =\begin{pmatrix}
        0\\
        1
    \end{pmatrix},\qquad \eta^{(2)} =-\begin{pmatrix}
        1\\
        0
    \end{pmatrix}.
\eeq
In the calculations presented below, a bar over the spinor denotes its Dirac adjoint. Additionally, in dealing with antiparticles, it is helpful to use the property
\begin{align}
    \begin{pmatrix}
        0&-1\\
        1&0
    \end{pmatrix}\varphi^{(r)}=\eta^{(r)} \quad \longrightarrow \quad -i  \sigma^2 \varphi^{(r)}=\eta^{(r)}.
\end{align}
Given the definitions of $\varphi^{(r)}$ and $\eta^{(r)}$, one can also show that an arbitrary two-by-two hermitian matrix $A$ satisfies the relations
\bel{}
    \varphi^{(r)\dagger} A \varphi^{(s)}&= A_{rs},
\eel
\bel{}
 \eta^{(s)\dagger} A \eta^{(r)}&= \varphi^{(s)\dagger} \sigma^2 A \sigma^2 \varphi^{(r)}=\varphi^{(s)\dagger} A^* \varphi^{(r)} = A^*_{sr} = A^\dagger_{rs}.
\eel

\section{II. Basic tensor expressions}
In the following, we drop the subscript $s$ in $X^{\pm}_s$ for clarity, i.e., $X^{\pm}$ below refers to the new definition of the spinor density matrix \EQn{XpmNEW2}. We recall that $X^{\pm}$ satisfies the commutation relation $[X^{\pm}, \slashed{p}] = [\gamma_5 \slashed{a}, \slashed{p}] =0$. 
\subsection{1. Baryon and particle currents}

We start from \EQ{eq:Nmu}
\bel{eq:NmuA} 
N^\mu(x) = \sum_{r=1}^2 \int \dd P  \, p^\mu \left[f^+_{rr}(x,p)-f^-_{rr}(x,p) \right] 
\eel
and use \EQSTWOn{fplusrsxp}{fminusrsxp} to find 
\begin{align}\begin{split}\label{eq:NmuA1} 
N^\mu(x) &=  \int \dd P  \, p^\mu \left[\trt\left(\frac{1}{2m} \,\ubarrp X^+ \usp\right)-\trt \left(- \frac{1}{2m} \, \vbarsp X^- \vrp \right) \right] \\
&= \frac{1}{2m}  \int \dd P  \, p^\mu \left[\trf\left[ X^+ (\slashed{p}+m)\right]+\trf \left[ X^- (\slashed{p}-m)\right] \right].
\end{split}\end{align}
Applying \EQ{XpmNEW3} together with the expansion given by \EQn{eq:expa}, we get
\begin{align}\begin{split}\label{eq:NmuA2} 
N^\mu(x) &= \frac{1}{2m}  \int \dd P  \, p^\mu \Bigg[\trf\left[ f_0^+(x,p)\left( \cosh\sqrt{-a^2} + \frac{\gamma_5 \slashed{a}}{\sqrt{-a^2}} \sinh\sqrt{-a^2}
\right) (\slashed{p}+m)\right]  \\
&\ \hspace{1cm} +\, \trf\left[f_0^-(x,p) \left( \cosh\sqrt{-a^2} + \frac{\gamma_5 \slashed{a}}{\sqrt{-a^2}} \sinh \sqrt{-a^2}
\right) (\slashed{p}-m)\right] \Bigg]  \\
&= 2  \int \dd P  \, p^\mu \left[  f_0^+(x,p)   - f_0^- (x,p)   \right]  \cosh \sqrt{-a^2}.
\end{split}\end{align}
The particle current defined in \EQ{eq:calNmu}, when computed, gives
\beq
    \mathcal{N}^\mu = 2\int \dd P  \, p^\mu \left[f^{+}_0(x,p)+f^{-}_0(x,p) \right] \cosh\sqrt{-a^2}.
\eeq
\subsection{2. Energy-momentum tensor}
Similarly to the prescription applied above, we take \EQ{eq:Tmunu} for the energy-momentum tensor,
\bel{eq:TmunuA} 
T^{\mu\nu}(x) = \sum_{r=1}^2 \int \dd P \, p^\mu p^\nu \left[f^+_{rr}(x,p)+f^-_{rr}(x,p) \right],
\eel
and rewrite it in terms of traces as
\begin{align}\begin{split}\label{eq:TmunuA2} 
T^{\mu\nu}(x) &=  \int \dd P  \, p^\mu p^\nu \left[\trt\left(\frac{1}{2m} \,\ubarrp X^+ \usp\right)+\trt \left(- \frac{1}{2m} \, \vbarsp X^- \vrp \right) \right] \\
&=\frac{1}{2m}  \int \dd P  \, p^\mu p^\nu \Big[\trf\left[X^+ (\slashed{p}+m)\right]-\trf \left[ X^- (\slashed{p}-m)\right] \Big]\\
&=\frac{1}{2m}  \int \dd P  \, p^\mu p^\nu \Bigg[\trf\left[ f_0^+(x,p)\left( \cosh\sqrt{-a^2} + \frac{\gamma_5 \slashed{a}}{\sqrt{-a^2}} \sinh \sqrt{-a^2}
\right) (\slashed{p}+m)\right] \\
&\ \hspace{1cm} -  \trf\left[f_0^-(x,p) \left( \cosh \sqrt{-a^2} + \frac{\gamma_5 \slashed{a}}{\sqrt{-a^2}} \sinh\sqrt{-a^2}
\right) (\slashed{p}-m)\right] \Bigg]  \\
&=2  \int \dd P  \, p^\mu p^\nu \left[  f_0^+(x,p)   + f_0^- (x,p)   \right]  \cosh\sqrt{-a^2}.
\end{split}\end{align}
\subsection{3. Spin tensor}
The spin tensor from \EQ{eq:Slmunu} reads
\bel{eq:SlmunuA} 
S^{\lambda, \mu\nu}(x) &=& \f{1}{2}\sum_{r,s=1}^2 \int \dd P \, p^\lambda  \left[\sigma^{+ \mu\nu}_{sr} f^+_{rs}(x,p)  + \sigma^{- \mu\nu}_{sr}f^-_{rs}(x,p) \right]
\eel
and can be explicitly computed utilizing again \EQSTWO{fplusrsxp}{fminusrsxp} as well as the definitions of $\sigma^{\pm \mu\nu}_{sr} (p)$. We obtain
 \bel{eq:SlmunuA2} 
S^{\lambda, \mu\nu}(x) &=& \frac{1}{8m^2}\sum_{r,s=1}^2 \int \dd P \, p^\lambda  \left( \left[\ubarsp \sigma^{\mu\nu} \urp \right] \left[\ubarrp X^+ \usp\right] + \left[ \vbarrp \sigma^{\mu\nu} \vsp \right] \left[ -  \, \vbarsp X^- \vrp\right] \right),
\eel
where by rearranging the terms, we can perform the summation leading to
\begin{align}\begin{split}\label{eq:SlmunuA3} 
S^{\lambda, \mu\nu}(x) &= \frac{1}{8m^2} \int \dd P \, p^\lambda \left( \trf \left[\sigma^{\mu\nu} (\slashed{p}+m) X^+ (\slashed{p}+m) \right] -
\trf \left[  \sigma^{\mu\nu} (\slashed{p}-m) X^- (\slashed{p}-m)
  \right]  \right) \\ 
&=  \frac{1}{8m^2} \int \dd P \, p^\lambda \left( \trf \left[\sigma^{\mu\nu}  X^+\, 2 m(\slashed{p}+m) \right] + \trf \left[  \sigma^{\mu\nu}  X^-\, 2 m(\slashed{p}-m)
  \right]  \right) \\
&=  \frac{1}{4m} \int \dd P \, p^\lambda \Bigg( \trf \left[\sigma^{\mu\nu}   f_0^+(x,p) \left( \cosh \sqrt{-a^2} + \frac{\gamma_5 \slashed{a}}{\sqrt{-a^2}} \sinh\sqrt{-a^2}
\right) (\slashed{p}+m) \right] \\
&\ \hspace{1cm} + \, \trf \left[  \sigma^{\mu\nu}  f_0^-(x,p) \left( \cosh \sqrt{-a^2} + \frac{\gamma_5 \slashed{a}}{\sqrt{-a^2}} \sinh\sqrt{-a^2}
\right) (\slashed{p}-m)
  \right]  \Bigg) \\
&=   \frac{1}{4m} \int \dd P \, p^\lambda \Bigg(  f_0^+(x,p)\,  \trf \left[\sigma^{\mu\nu} \frac{\gamma_5 \slashed{a} \slashed{p}}{\sqrt{-a^2}} \sinh\sqrt{-a^2}\right]  + f_0^-(x,p)\,  \trf \left[ \sigma^{\mu\nu} 
 \frac{\gamma_5 \slashed{a} \slashed{p}}{\sqrt{-a^2}} \sinh\sqrt{-a^2}
 \right]  \Bigg)\\
&=  \frac{1}{4m} \int \dd P \, p^\lambda\,  \left[  f_0^+(x,p) + f_0^-(x,p)  \right] \frac{\sinh\sqrt{-a^2}}{\sqrt{-a^2}} \trf \left[ \sigma^{\mu\nu} 
\gamma_5 \slashed{a} \slashed{p}
 \right] \\
 &= \frac{1}{m} \int \dd P \, p^\lambda\, 
\left[ f_0^+(x,p) + f_0^-(x,p)  \right] \frac{\sinh\sqrt{-a^2}}{\sqrt{-a^2}}  {\epsilon^{\mu \nu}}_{\rho \sigma} a^\rho p^\sigma.
\end{split}\end{align}
Above, the trace was computed using the expression
 \bel{eq:TrGamma}
\trf\left(\gamma ^{\mu }\gamma ^{\nu }\gamma ^{\rho }\gamma ^{\sigma }\gamma ^{5}\right)=-4i\epsilon ^{\mu \nu \rho \sigma }.
 \eel

\section{III. Thermodynamic identities}

\subsection{1. Entropy current}
We start from the definition of the entropy current given by \EQ{eq:Smu},
\begin{equation}
     S^\mu = - \frac{1}{2} \int \dd P p^\mu \Big\{\trf\left[X^+ \left(\ln X^+ - 1\right)\right] + \trf\left[X^- \left(\ln X^- - 1\right)\right]\Big\},
\end{equation}
which can be rewritten using the form of $X^\pm$ defined in \EQ{XpmNEW2}, 
\begin{align}\begin{split}
  X^\pm \ln X^\pm &= f^{ \pm}_0 \exp\left( \gamma_5 \slashed{a}\right) \ln\Big[f^{ \pm}_0 \exp\left( \gamma_5 \slashed{a}\right)\Big] = f^{ \pm}_0 \exp\left( \gamma_5 \slashed{a}\right) \Big(\ln f^{ \pm}_0 + \gamma_5 \slashed{a}\Big) \\
    &= f^{ \pm}_0 \exp\left( \gamma_5 \slashed{a}\right) \ln f^{ \pm}_0  + f^{ \pm}_0 \exp\left( \gamma_5 \slashed{a}\right) \gamma_5 \slashed{a}.
\end{split}\end{align}
Next, we determine the traces
\begin{align}\begin{split}\label{Term1}
    \trf \Big[X^\pm \ln X^\pm\Big] &= \trf \Big[f^{ \pm}_0 \ln f^{ \pm}_0 \exp\left( \gamma_5 \slashed{a}\right) + f^{ \pm}_0 \exp\left( \gamma_5 \slashed{a}\right) \gamma_5 \slashed{a}\Big] \\
    &= \left(f^{ \pm}_0 \ln f^{ \pm}_0\right) \cosh\sqrt{- a^2}\,\, \trf \Big[1 + \frac{\gamma_5 \slashed{a}}{\sqrt{-a^2}} \tanh\sqrt{-a^2}\Big] \\
    &\
\hspace{1cm}  +\, f^{ \pm}_0 \cosh\sqrt{- a^2}\,\, \trf \left[\left(1 + \frac{\gamma_5 \slashed{a}}{\sqrt{-a^2}} \tanh\sqrt{-a^2}\right)\gamma_5 \slashed{a}\right]\\
    &= 4 \left(f^{ \pm}_0 \ln f^{ \pm}_0\right) \cosh\sqrt{- a^2} + f^{ \pm}_0\, \frac{\sinh\sqrt{- a^2}}{\sqrt{- a^2}}\, \trf \left[\gamma_5 \gamma_\mu \gamma_5 \gamma_\nu\right] a^\mu a^\nu \\
&= 4 \left(f^{ \pm}_0 \ln f^{ \pm}_0\right) \cosh\sqrt{- a^2} - 4 f^{ \pm}_0\,\frac{\sinh\sqrt{- a^2}}{\sqrt{- a^2}}  g_{\mu\nu} a^\mu a^\nu \\
&=  4 \left[\left(f^{ \pm}_0 \ln f^{ \pm}_0\right) \cosh\sqrt{- a^2} - f^{ \pm}_0\, \frac{\sinh\sqrt{- a^2}}{\sqrt{- a^2}} a^2\right]
\end{split}\end{align}
and
\begin{align}\begin{split}\label{Term2}
    \trf \Big[X^+ + X^-\Big] &= \trf \Big[f^{ +}_0 \exp\left(\gamma_5 \slashed{a}\right) + f^{ -}_0 \exp\left( \gamma_5 \slashed{a}\right)\Big]  \\
     &= \trf \Bigg[f^{ +}_0 \cosh\sqrt{-a^2} \left(1 + \frac{\gamma_5 \slashed{a}}{\sqrt{- a^2}} \tanh\sqrt{-a^2}\right) + f^{ -}_0 \cosh\sqrt{-a^2} \left(1 + \frac{\gamma_5 \slashed{a}}{\sqrt{- a^2}} \tanh\sqrt{-a^2}\right) \Bigg]\\
     &= 4 \left(f^{ +}_0 + f^{ -}_0\right) \cosh\sqrt{-a^2}. 
\end{split}\end{align}
Summing up the two terms from \EQS{Term1} and \EQn{Term2}, and putting them back into \EQ{eq:Smu}, we obtain
\bel{eq:SmuA1}
    S^\mu \!= \!- 2 \int \dd P\, p^\mu \left[\Big(\!\left[f^{ +}_0 (\xi-p_\mu\beta^\mu) \!-\! f^{ +}_0\right] \!-\! \left[f^{ -}_0 (\xi+p_\mu\beta^\mu) \!+\! f^{ -}_0\right]\! \Big) \cosh\sqrt{- a^2} - \left(f^{ +}_0 \!+\! f^{ -}_0\right) \! \frac{\sinh\sqrt{- a^2}}{\sqrt{- a^2}}\, a^2\right]. 
\eel
The expressions obtained in \EQS{eq:NmuA2}, \EQn{eq:TmunuA2}, and \EQn{eq:SlmunuA3}, when taken together with the above formula, satisfy \EQ{eq:H}.
\subsection{2. Gibbs-Duhem relation}
The differential of the particle current $\mathcal{N}^\mu$ is
\begin{align}\begin{split}
    \dd\,\mathcal{N}^\mu &= 2\int \dd P  \, p^\mu \sinh\sqrt{-a^2}\,\, \dd\left(\sqrt{-a^2}\right) \left(f^{ +}_0+f^{ -}_0 \right)+2\int \dd P  \, p^\mu \left(\dd\,f^{ +}_0+\dd\,f^{ -}_0 \right) \cosh\sqrt{-a^2}  \\
    &= 2\int \dd P  \, p^\mu \sinh \sqrt{-a^2} \dd\left(\sqrt{-a^2}\right) \left(f^{ +}_0+f^{ -}_0 \right)+2\dd\xi\int \dd P  \, p^\mu \left(f^{ +}_0-f^{ -}_0 \right) \cosh\sqrt{-a^2}  \\
&\     \hspace{1cm} - 2\int \dd P  \, p^\mu p^\nu \dd\beta_\nu \left(f^{ +}_0+f^{ -}_0 \right) \cosh\sqrt{-a^2}  \\
    &= -2\int \dd P  \, p^\mu \frac{\sinh\sqrt{-a^2} }{\sqrt{-a^2}}a_\rho \dd a^\rho \left(f^{ +}_0+f^{ -}_0 \right) + \dd\xi N^\mu - T^{\mu\nu}\dd\beta_{\nu} \\
    &= - 2\int \dd P  \, p^\mu  \frac{\sinh\sqrt{-a^2}}{\sqrt{-a^2}}a_\rho \left(-\frac{\epsilon^{\rho\nu\alpha\beta}p_\nu \dd\omega_{\alpha\beta}}{4m}\right) \left(f^{ +}_0+f^{ -}_0 \right) + \dd\xi N^\mu - T^{\mu\nu}\dd\beta_{\nu} \\
    &=  \dd\omega_{\alpha\beta} \int \dd P  \, p^\mu   \epsilon^{\alpha\beta\rho\nu}
    \frac{p_\nu  a_\rho}{2m} \frac{\sinh\sqrt{-a^2}}{\sqrt{-a^2}}  \left(f^{ +}_0+f^{ -}_0 \right) + \dd \xi N^\mu - T^{\mu\nu}\dd \beta_{\nu} \\
    &= \frac{1}{2}S^{\mu, \alpha \beta} \dd \omega_{\alpha\beta}+ \dd \xi N^\mu - T^{\mu\nu}\dd\beta_{\nu}.
\end{split}\end{align}
Thus, the Gibbs-Duhem relation is satisfied.
\subsection{3. First law of thermodynamics}

With the agreement of the microscopic and macroscopic definitions of the entropy current and the fulfillment of the Gibbs-Duhem relation, the first law of thermodynamics emerges naturally.

\section{IV. Connection to classical-spin approach}

\subsection{1. Classical spin variables and integration measure}

In the classical approach to spin~\CITn{Mathisson:1937zz,2010GReGr..42.1011M}, the tensor $s^{\alpha\beta}$ is defined as
\bel{eq:sab}
s^{\alpha\beta} = \f{1}{m} \epsUabgd p_\gamma s_\delta,
\eel
which by definition is antisymmetric and orthogonal to the particle four-momentum $p^\beta$. One can invert the relation~\EQn{eq:sab} and express $s^\alpha$ in terms of the second-rank antisymmetric tensor given above, 
\bel{eq:sa}
s^{\alpha} = \f{1}{2m} \,\epsUabgd p_\beta s_{\gamma \delta}.
\eel
In the PRF, the time component of $s^\alpha$ vanishes, leaving only the space component given by 
$s^\alpha = (0,\sv_*)$. The magnitude of the spacelike component is set to the square root of the eigenvalue of the Casimir operator $\spin^2$, which equals $\nicefrac{3}{4}$ for spin-$\onehalf$ particles.

To apply the principles of kinetic theory to particles with spin, one must include the quantity $s^\alpha$ as a parameter of the distribution function. The distribution function in the standard spinless case is written as $f(x,\pv)$, which is used interchangeably with $f(x,p)$, bearing in mind that the energy is on the mass shell, i.e., $p^0= E_{\rm p}= \sqrt{m^2 + \pv^2}$. While extending this approach to include spin, we consider functions of the type $f(x,p,s)$, which should be considered together with the following integration measures~\CITn{Florkowski:2018fap}
\bel{eq:dP_dS}
\dd P = \f{\dd^3p}{(2\pi)^3 E_{\rm p}}, \qquad \dd S = \f{m}{\pi \spin}  \, \dd^4 s \, \delta(s \cdot s + \spin^2) \, \delta(p \cdot s).
\eel
The spin measure includes two delta functions that restrict the domain of the integration only to values that satisfy the normalization condition and obey the orthogonality to the four-momentum. The prefactor appearing in this measure is chosen such that
\bel{eq:dS2}
\int \dd S = 2.
\eel
We additionally provide some useful integrals in the spin space~\CITn{Florkowski:2018fap}:
\bel{eq:dSs} 
\int \dd S \,s_{\alpha} &=& 0, \qquad
\int \dd S \,s_\alpha s_\beta = 
-\f{2 }{3} \spin^2\LB g_{\alpha \beta} - \frac{p_\alpha p_\beta}{m^2} \RB. \label{eq:dSss}
\eel
It is possible to show that all integrals of a product of an odd number of $s$ vanish.

\subsection{2. Quadratic corrections due to the spin polarization tensor}

The presence of the spin polarization tensor $\omega$ introduces quadratic corrections to the baryon current and the energy-momentum tensor. The easiest way to study such corrections is to analyze the modification of the spin integral that in the two cases has the same form
\bel{class_cor1}
1 + \f{1}{16} \int \dd S \, \omega_{\alpha \beta} s^{\alpha\beta} \, \omega_{\rho \sigma}  s^{\rho \sigma}.
\eel
In the quantum approach, the corrections to the baryon current and the energy-momentum tensor emerge from the expansion of the function $\cosh\sqrt{-a^2}$, namely
\bel{quant_cor1}
\cosh\sqrt{-a^2} = 1 - \f{1}{2} a^2 + \mathcal{O}(a^4).
\eel
We arrive at the above equation by expressing $s^{\alpha \beta}$ in \EQn{class_cor1} in terms of $s^\delta$, using additionally \EQ{eq:dSs}.

The spin tensor contains only odd powers of the spin polarization tensor. Therefore, at the lowest nontrivial order, we can restrict our considerations to terms linear in $\omega$ only. In this case, our new expression for $S^{\lambda, \mu\nu}$ (using $X^\pm_s)$ agrees with the old expression (using $X^\pm$), as pointed out above \EQ{eq:classEQ} in the main text. Since the old formula is known to agree with the classical result, our new expression agrees as well. This consistency can also be verified by a straightforward calculation.

\section{V. Causality of spin hydrodynamics}

Starting from \EQ{eq:MlAB}, with the generating function $\chi$ defined by \EQ{eq:chi}, one can prove that 
\begin{align}
    M^\lambda \equiv M^{\lambda AB} Z_A Z_B = \left( Z\frac{\partial}{\partial\xi} - Z_{\mu}\frac{\partial}{\partial\beta_{\mu}} + \frac{1}{2} Z_{\mu\nu} \frac{\partial}{\partial \omega_{\mu\nu}}\right)^2 {\cal N}^\lambda. \label{M-ZZ}
\end{align}
Here the vector $Z_A$ has components $Z_A = (Z, Z_\mu, Z_{\mu\nu})$ and the factor $\onehalf$ eliminates double counting in the summation over the indices $\mu \nu$.  After straightforward calculations, we obtain 
\begin{align}\begin{split}
M^{\lambda}
    &= \frac{\cosh\sqrt{-a^2}}{{2}} \Bigg\{4 \int \dd P\, p^\lambda \Big[(Z + Z\cdot p) f_0^+ (x,p) + (- Z + Z\cdot p) f_0^- (x,p)\Big] \frac{\tanh\sqrt{-a^2}}{\sqrt{- a^2}} \, \frac{a\cdot \widetilde{Z}\cdot p}{m}\\
    &\ \hspace{1cm}  - \int \dd P\, p^\lambda \Big[f_0^+ (x,p) + f_0^- (x,p)\Big] \frac{(a\cdot \widetilde{Z}\cdot p)^2}{m^2\, a^2}   \\
    &\ \hspace{1cm}  + 4 \int \dd P\, p^\lambda \left[\left(Z + Z\cdot p \right)^2 f_0^+ (x,p) + (- Z + Z\cdot p )^2 f_0^- (x,p) \right] \Bigg\} \\
    &\ \hspace{1cm}  - \frac{\sinh{\sqrt{-a^2}}}{2} \Bigg[\int \dd P\, p^\lambda \Big[f_0^+ (x,p) + f_0^- (x,p)\Big] \frac{(a\cdot \widetilde{Z}\cdot p)^2}{m^2\, \left(-a^2\right)^{3/2}}  \\
    &\ \hspace{1cm}  + \int \dd P\, p^\lambda \Big[f_0^+ (x,p) + f_0^- (x,p)\Big] \frac{1}{m^2\,\sqrt{- a^2}}  \left(\widetilde{Z}_{\alpha\gamma} p^\gamma\right) \left(\widetilde{Z}^{\alpha\beta} p_\beta\right)\Bigg],
    \end{split}
\end{align}
where $\widetilde{Z}^{\mu\nu} = - \widetilde{Z}^{\nu\mu} = (1/2) \varepsilon^{\mu\nu\alpha\beta} Z_{\alpha\beta}$ is the tensor dual to $Z_{\alpha\beta}$, and $a\cdot \widetilde{Z}\cdot p = a_\mu \widetilde{Z}^{\mu\nu} p_\nu$. The last equation has the structure
\begin{align}
    M^\lambda &= \int \dd P \,p^\lambda \, w,
\end{align}
where $w$ is a scalar function that can be calculated in the PRF, namely
\begin{align}
\begin{split}
    w &= \frac{\cosh\sqrt{\av_*^{\,2}}}{2} \left(4\gamma^2_+ + \left(\nv \cdot \alphav \right)^2 - 4 \gamma_+ \nv \cdot \alphav  \tanh\sqrt{\av_*^{\,2}} \right) f_0^+ (x,p)\\
     &\ \hspace{1cm} + \frac{\cosh\sqrt{\av_*^{\,2}}}{2} 
    \left(4 \gamma^2_- + \left(\nv \cdot \alphav\right)^2 - 4 
    \gamma_- \nv \cdot \alphav \tanh\sqrt{\av_*^{\,2}} \right) f_0^- (x,p) \\
     &\ \hspace{1cm} + \frac{\sinh{\sqrt{\av_*^{\,2}}}}{2 \sqrt{\av_*^{\,2}}} \Big( \alphav \cdot \alphav - \left(\nv \cdot \alphav \right)^2\!\!\Big) \Big(f_0^+ (x,p) + f_0^- (x,p)\Big). \label{W*-def}
    \end{split}
\end{align}
Here the asterisk denotes the quantities in PRF, in particular, we have $a^2 = a^2_* = -\av^2_*$. We have also introduced the notation: $\gamma_\pm = \left(\pm Z + Z\cdot p\right)^*$, $\nv = \av_*/\sqrt{\av_*^{\,2}}$ (with $\nv^2=1$), and $\alpha^i = \widetilde{Z}^{i0}$. Since $\tanh\sqrt{\av_*^{\,2}} \in [-1,1]$, we can conclude that the first two lines in~\EQn{W*-def} are positive expressions. One can easily check that the third line of~\EQn{W*-def} is positive as well. 

\section{VI. Selection rules for the polarization tensor}

Practically, all studies of spin polarization that use a concept of the spin polarization tensor $\omega$ (or a spin chemical potential $\Omega = \omega T$) assume that the components of $\omega$ are small. This treatment is supported by the fact that the experimentally measured effects are in fact small, typically below 1\%. Therefore, most of the time one includes only the leading (first-order) terms in $\omega$. The inclusion of higher-order terms has become only a topic of more recent works~\cite{Florkowski:2024bfw, Drogosz:2024gzv, Drogosz:2025ihp, Abboud:2025qtg, Drogosz:2025iyr}, where the expansion in $\omega$ is studied not only because of the practical reasons but also because of the interest in the internal consistency of the constructed formalism. In this section, we first comment on the most common approximation used in the literature, and then we switch to a discussion of more recent results that incorporate all orders in $\omega$.

\subsection{1. Traditional approach}

As we discuss in the main text, in most of the analyzes performed so far, the crucial assumption made to deal with well-defined polarization was that the expression $\sqrt{ \bv_\ast \cdot \bv_\ast \!-\! \ev_\ast \cdot \ev_\ast }$ is small. In this case, the mean polarization is directly expressed by the vector $ \bv_\ast$ whose explicit formula reads~\CITn{Florkowski:2017dyn} 
\bel{eq:bstar1}
 \bv_\ast &=& \f{1}{m} \left(  E_{\rm p} \, \bv - \pv \times \ev - \f{\pv \cdot \bv}{E_{\rm p} + m} \pv \right).
\eel
It is worth emphasizing that an assumption of small $\ev$ and $\bv$ is insufficient to ensure a small $\bv_\ast$. The first term in \EQn{eq:bstar1} includes the Lorentz gamma factor $E_{\rm p}/m$, which may be very large for relativistic particles and compensate for the smallness of $\bv$. Thus, the assumption about the smallness of $\bv_\ast$ excludes from the analysis some portion of relativistic particles.

\subsection{2. New developments}

The allowed region for the Lagrange multipliers appearing in our approach has recently been studied in more detail in~\cite{Drogosz:2025ihp}. The main result of that study is the formula that specifies the range of the electric and magnetic components of $\omega$, 
\begin{eqnarray}\label{eq:criterionW}
\f{1}{2} \sqrt{ {\bvp}^2 + {\evp}^2 + 2 |\evp \times \bvp|} < \f{m}{T}.
\label{eq:condition1}
\end{eqnarray}
The primes denote vectors in the fluid rest frame, where $u^\mu=(1,0,0,0)$. If the cross product vanishes, \EQ{eq:condition1} is reduced to that found in~\CITn{Abboud:2025qtg}. It is important to stress that \EQn{eq:condition1} holds for all real systems that have finite densities of energy, linear and orbital momentum, and spin.

As explained in~\cite{Drogosz:2025ihp}, in practical calculations one may use a more restrictive formula
\begin{equation}
\omega_{\rm max} \, \sqrt{ \f{1+v}{1-v} } \,
< \kappa \, \frac{m}{T} \,.  
\label{eq:condition2}
\end{equation}
Here $\omega_{\rm max}$ is the magnitude  (absolute value) of the largest component of the spin polarization tensor $\omega_{\alpha\beta}$ in the LAB frame, while $\kappa$~is a numerical factor of the order of unity, \mbox{$\kappa = 2/\left(2+\sqrt{2}\right)$.}  The proof of~\EQn{eq:condition2} is given below, after we make a few important comments:
\begin{itemize}
\item[i)] At the top RHIC energies, the radial flow of matter usually does not exceed half of the speed of light, while the polarization effects (quantified by the magnitude of $\omega_{\rm max}$) are at the level of a fraction of a percent. Moreover, our model is expected to work at the late stages of the collisions, where the temperatures are close to $0.15$~GeV, and the commonly considered masses are~\CITn{Becattini:2021iol,Fu:2021pok}: $m=1.1$~GeV (the~$\Lambda$ hyperon mass) or $m=0.5$~GeV (the constituent strange quark mass). Using these numbers in \EQ{eq:condition2}, we observe that this condition is fulfilled very well. At lower energies, we deal with smaller values of flow and lower temperatures, which allow to fulfill~\EQn{eq:condition2} even with much larger values of~$\omega_{\rm max}$.

\item[ii)] The condition~\EQn{eq:condition2} is satisfactory to have all thermodynamic variables well defined. We checked this property by numerical calculations of different variables, such as the baryon or energy density. Inequality~\EQn{eq:condition2} can be relaxed (larger values of $\kappa$ can be allowed), if the fields $\bv$ and $\ev$ exhibit certain symmetries. For example, if one of them is zero, $\kappa = 2/\sqrt{3}$.

\item[iii)] The factor $\sqrt{(1+v)/(1-v)}$ in~\EQ{eq:condition} can be naturally interpreted as the longitudinal relativistic Doppler effect that increases ``frequency''  $\omega_{\rm max}$ to that ``seen'' in the local fluid rest frame. 

\item[iv)] Last but not least, we stress that~\EQn{eq:condition2} includes only hydrodynamic variables and is different from the previously used conditions that mix hydrodynamic variables with single-particle properties (as discussed in the subsection above).
    
\end{itemize}

Let us now turn to the proof of~\EQn{eq:condition2}. The new definition of the Wigner function includes an exponential factor that may possibly diverge for large values of the three-momentum $|\pv|$. It has the form
\begin{eqnarray}
\exp \left(
-\frac{p_\mu u^\mu}{T}
+ \frac{1}{2 m} \sqrt{-
{\tilde \omega}_{\mu \alpha } p^\alpha 
{\tilde \omega}^\mu_\beta p^\beta
}  
\right) = 
\exp \left(
- \, \frac{p_\mu u^\mu}{T} \right) \exp \left(
 \, \frac{1}{2 m} \sqrt{-
{\tilde \omega}_{\mu \alpha } p^\alpha 
{\tilde \omega}^\mu_\beta p^\beta
}  
\right).
\end{eqnarray}
The divergence appears if the damping by the temperature part is not sufficient to suppress the second term. Since this expression is Lorentz-invariant, we can first consider the LAB frame where $u^\mu = \gamma (1, \vv )$
and
\bel{omdeb}
\omnLD = 
\begin{bmatrix}
	0       &  b^1 & b^2 & b^3 \\
	-b^1  &  0    & e^3 & -e^2 \\
	-b^2  &  -e^3 & 0 & e^1 \\
	-b^3  & e^2 & -e^1 & 0
\end{bmatrix}.
\eel
We note that to switch from $\omnL$ to the dual tensor $\omnLD$, one replaces $\ev$ by $\bv$ and $\bv$ by~$-\ev$. 

In the next step, it is convenient to transform $u$ and $p$ to the local rest frame of the fluid element; hence, we write
\begin{eqnarray}
u^\mu = \Lambda^\mu_{\,\,\,\alpha^\prime} u^{\alpha^\prime}, \qquad   
p^\mu = \Lambda^\mu_{\,\,\,\alpha^\prime}
p^{\alpha^\prime}, \qquad 
p_\mu = \Lambda_\mu^{\,\,\,\alpha^\prime}
p_{\alpha^\prime},
\end{eqnarray}
where $\Lambda^\mu_{\,\,\,\alpha^\prime} = \gamma \lambda^\mu_{\,\,\,\alpha^\prime}$, with
\bel{Lambda}
\lambda^\mu_{\,\,\,\alpha^\prime} = 
\begin{bmatrix}
1 & \quad \frac{\sqrt{\gamma^2-1}}{\gamma} {\hat v}^1 & \quad \frac{\sqrt{\gamma^2-1}}{\gamma}{\hat v}^2 & \quad \frac{\sqrt{\gamma^2-1}}{\gamma} {\hat v}^3 \\[0.5em]
\frac{\sqrt{\gamma^2-1}}{\gamma} {\hat v}^1 & \quad  \f{1}{\gamma} + \f{\gamma-1}{\gamma} {\hat v}^1 {\hat v}^1 
& \quad  \f{\gamma-1}{\gamma} {\hat v}^1 {\hat v}^2  & \quad \f{\gamma-1}{\gamma} {\hat v}^1 {\hat v}^3 \\[0.5em]
\frac{\sqrt{\gamma^2-1}}{\gamma} {\hat v}^2 & \quad  \f{\gamma-1}{\gamma} {\hat v}^2 {\hat v}^1 
& \quad  \f{1}{\gamma} + \f{\gamma-1}{\gamma} {\hat v}^2 {\hat v}^2 & \quad  \f{\gamma-1}{\gamma} {\hat v}^2 {\hat v}^3 \\[0.5em]
\frac{\sqrt{\gamma^2-1}}{\gamma} {\hat v}^3 & \quad  \f{\gamma-1}{\gamma} {\hat v}^3 {\hat v}^1 
& \quad \f{\gamma-1}{\gamma} {\hat v}^3 {\hat v}^2 & \quad  \f{1}{\gamma} + \f{\gamma-1}{\gamma} {\hat v}^3 {\hat v}^3
\end{bmatrix}.
\eel
Here ${\hat \vv}$ is the unit three-vector parallel to $\vv$, ${\hat \vv} = \vv/|\vv|\equiv\vv/v$. Since we extracted $\gamma$ from $\Lambda^\mu_{\,\,\,\alpha^\prime}$, all elements of the matrix $\lambda^\mu_{\,\,\,\alpha^\prime}$ are finite and bounded ($1/\gamma < 1$, \ $\sqrt{\gamma^2-1}/\gamma < 1$, \  $(\gamma-1)/\gamma < 1$, \  $|{\hat v}^i| < 1$). We can perform a similar transformation to the matrix $\omnLD$ by extracting the modulus of its largest element, namely,
\bel{omdeb2}
\omnLD = 
\begin{bmatrix}
	0       &  b^1 & b^2 & b^3 \\
	-b^1  &  0    & e^3 & -e^2 \\
	-b^2  &  -e^3 & 0 & e^1 \\
	-b^3  & e^2 & -e^1 & 0
\end{bmatrix} 
= \omega_{\rm max}
\begin{bmatrix}
0  &  {\hat b}^1 & {\hat b}^2 & {\hat b}^3 \\
-{\hat b}^1 &  0    & {\hat e}^3 & -{\hat e}^2 \\
-{\hat b}^2 & -{\hat e}^3 & 0 & {\hat e}^1 \\
-{\hat b}^3 & {\hat e}^2 & -{\hat e}^1 & 0
\end{bmatrix}
\equiv \omega_{\rm max} \, {\hat \omega}_{\mu\nu},
\eel
where ${\hat \bv} = \bv/\omega_{\rm max}$ and ${\hat \ev} = \ev/\omega_{\rm max}$ (so we have $|{\hat b}^i| < 1$ and $|{\hat e}^i| < 1$). Thus, similarly to $\lambda^\mu_{\,\,\,\alpha^\prime}$, the matrix ${\hat \omega}_{\mu\nu}$ contains bounded elements.

Using the notation introduced above, we can rewrite the argument of the exponential function as 
\begin{eqnarray}
-\frac{E_{\rm p^\prime}}{T}
+ \frac{\gamma \, \omega_{\rm max} }{2 m} \sqrt{-
{\hat \omega}_{\mu \alpha } \lambda^\alpha_{\alpha^\prime} p^{\alpha^\prime}
{\hat \omega}^\mu_\beta \lambda^\beta_{\beta^\prime} p^{\beta^\prime} }   .
\label{eq:arg1}
\end{eqnarray}
Here $E_{\rm p^\prime} = \sqrt{ m^2+\pv^{\prime^2}}$ and $p^{\mu^\prime} = (E_{\rm p^\prime}, \pv^\prime)$. In the limit $|\pv^\prime| \to \infty$, we can use the approximations: $E_{\rm p^\prime} = |\pv^\prime|$ and $p^{\mu^\prime} = |\pv^\prime| (1, {\hat \pv}^\prime)$, where ${\hat \pv}^\prime = \pv^\prime/ |\pv^\prime|$. Hence, for very large values of $|\pv^\prime|$, which determine the convergence of thermodynamic integrals, we can rewrite~\EQn{eq:arg1} as
\begin{eqnarray}
\left( 
-\frac{1}{T}
+ \frac{\gamma \, \omega_{\rm max} }{2 m} \sqrt{-
{\hat \omega}_{\mu \alpha } \lambda^\alpha_{\alpha^\prime} 
{\hat p}^{\alpha^\prime}
{\hat \omega}^\mu_\beta \lambda^\beta_{\beta^\prime} 
{\hat p}^{\beta^\prime} }   
\right) |\pv^\prime|,
\label{eq:arg2}
\end{eqnarray}
where we introduced the notation ${\hat p}^{\,\mu^\prime} = (1, {\hat \pv}^\prime)$.

Since all the elements entering the square root $\mathcal{R} \equiv \sqrt{-
{\hat \omega}_{\mu \alpha } \lambda^\alpha_{\alpha^\prime} 
{\hat p}^{\alpha^\prime}
{\hat \omega}^\mu_\beta \lambda^\beta_{\beta^\prime} 
{\hat p}^{\beta^\prime} }$ are bounded, the expression is bounded from above. To find its upper bound, we combined numerical and analytical methods. We performed a 10-dimensional Monte Carlo maximization (with respect to $\hat b^i$, $\hat e^i$, and four angles defining the unit vectors $\hat \vv$ and $\hat \pv$) with iterative shrinkage of the probed regions of the parameter space that were found to contain the highest values of $\mathcal R$. As the regions appeared to converge toward certain well-defined points, we constructed a surmise that those points indeed correspond to the maxima, and we confirmed it by calculating the partial derivatives of $\mathcal R$ at the given points. The details will be presented in a separate work.
We found that the maximum of $\mathcal R$ is equal to $(2+\sqrt{2})(1+v)$. Consequently, a sufficient condition for convergence of thermodynamic integrals is 
\begin{eqnarray}
-\frac{1}{T}
+ \frac{\gamma \, \omega_{\rm max} }{2 m}  
(2+\sqrt{2})(1+v) 
=
-\frac{1}{T}
+ \frac{ \omega_{\rm max} }{m}  
\f{(2+\sqrt{2})}{2} \sqrt{\f{1+v}{1-v} }< 0
\label{eq:arg3}
\end{eqnarray}
or
\begin{eqnarray}
\omega_{\rm max}  \sqrt{\f{1+v}{1-v} }
\, < \, \f{2}{2+\sqrt{2}} \, \frac{m}{T} 
\, \equiv \, \kappa \frac{m}{T}.
\label{eq:arg4}
\end{eqnarray}
This expression corresponds to~\EQn{eq:condition2}. 

\end{widetext}

\end{document}